\documentclass{osa-article}

\usepackage[utf8]{inputenc}
\usepackage{amsmath}
\usepackage{tikz}           
\usetikzlibrary{patterns}   
\usepackage{bm}             
\usepackage{amsfonts}       
\usepackage{hyperref}       
\usepackage{graphicx}
\graphicspath{{./figures/}}
\usepackage{subcaption} 
\usepackage{algorithm} 
\usepackage[noend]{algpseudocode} 
\usepackage{booktabs} 
\usepackage{adjustbox} 
\usepackage{mathtools} 
\usetikzlibrary{shapes,arrows,fit,backgrounds}
\usepackage{titling}  

\DeclareMathOperator*{\argmin}{argmin}
\DeclareMathOperator*{\median}{median}
\renewcommand{\vec}[1]{\bm{#1}}
\DeclarePairedDelimiter\abs{\lvert}{\rvert}

\newcommand{\areal}{{\rho_A}}
\newcommand{\MAC}{\xi}
\newcommand{\photons}{{q}}
\newcommand{\gain}{{\vec{c}}}
\newcommand{\dens}{{\vec{\rho}}}
\renewcommand{\flat}{{\vec{f}}}
\newcommand{\dir}{{\vec{d}}}
\newcommand{\sca}{{\vec{s}}}
\newcommand{\tot}{{\vec{t}}}
\renewcommand{\ker}{{\vec{k}}}

\newcommand{\model}{\mathcal{A}}

\makeatletter
\providecommand*{\input@path}{}
\g@addto@macro\input@path{{figures/}}
\makeatother

\journal{oe}

\begin{document}


\title{Local Models for Scatter Estimation and Descattering in Polyenergetic X-Ray Tomography}

\author{Michael~T.~McCann,\authormark{1,2,*}
Marc~L.~Klasky,\authormark{2}
Jennifer~L.~Schei,\authormark{3} 
and Saiprasad~Ravishankar\authormark{1,4}}

\address{
	    \authormark{1} Department of Computational Mathematics, Science and Engineering, Michigan State University, East Lansing, MI, 48824, USA (previously)\\
	    \authormark{2} Theoretical Division, Los Alamos National Laboratory, Los Alamos, NM, 87545, USA\\
		\authormark{3} Physics Division, Los Alamos National Laboratory, Los Alamos, NM, 87545, USA\\
		\authormark{4} Department of Biomedical Engineering, Michigan State University, East Lansing, MI, 48824 USA 
}

\email{\authormark{*}mccann13@msu.edu} 

	\begin{abstract}
	We propose a new modeling approach for scatter estimation and descattering in polyenergetic X-ray computed tomography (CT) based on fitting models to local neighborhoods of 
	a training set.
	X-ray CT is widely used in medical and industrial applications.
	X-ray scatter, if not accounted for during reconstruction, 
	creates a loss of contrast in CT reconstructions and introduces severe artifacts including cupping, shading, and streaks.
	Even when these qualitative artifacts are not apparent,
	scatter can pose a major obstacle in obtaining quantitatively accurate  reconstructions.
	Our approach to estimating scatter is, first, to generate a training set of 2D radiographs
	with and without scatter using particle transport simulation software.
	To estimate scatter for a new radiograph, 
	we adaptively fit a scatter model to a small subset of the training data
	containing the radiographs most similar to it.
	We compared local and global (fit on full data sets) versions of several X-ray scatter models,
	including two from the recent literature,
	as well as a recent deep learning-based scatter model,
	in the context of descattering and quantitative density reconstruction
	of simulated, spherically symmetrical, single-material objects
	comprising shells of various densities.
	Our results show
	that, when applied locally,
	even simple models provide state-of-the-art descattering,
	reducing the
	error in density reconstruction due to scatter by more than half.
	\end{abstract}

\section{Introduction} \label{sec:intro}
In the event of a possible nuclear threat, the United States government employs an Emergency Response Team to determine the proper course of action.
A crucial step in assessing the threat is a determination of the object characteristics,
and radiographic examination of the object is an important part of this process.
While tomographic reconstruction techniques
such as filtered backprojection or model-based iterative reconstruction (MBIR)~\cite{ravishankar_image_2020}
are available,
great difficulty remains in determining the material composition of the object due to scattered radiation, spectral effects, and issues associated with the detector. 
Similar challenges exist in the context of baggage screening
and nondestructive testing.

This work focuses on addressing scattered radiation in service of performing quantitative radiographic reconstructions.
In general, scatter creates loss of contrast impairing the ability to find sharp features as well as introducing severe image artifacts, e.g., cupping, shading, streaks, etc.
Consequently, in applications in which a quantitative density reconstruction is desired, 
scatter must be mitigated in hardware
(e.g., by increasing the distance between the object and the image plane, 
or through use of a Bucky-Potter grid)
and/or removed in software during reconstruction.

Scatter is a complex phenomenon with many possible types of individual radiation-matter interactions
(as X-ray photons pass through an object),
e.g., Compton scatter, Rayliegh scatter, pair production, and the photoelectric effect~\cite{cohentannoudji_atom_1998}. 
The total amount of scatter along a ray is simply proportional to the probability of interaction~\cite{cohentannoudji_atom_1998},
but the difficulty lies in computing the spatial distribution of scatter reaching the detector.
The transport of the scattered radiation is in a direction away from the direction of the initial scattering interaction. 
Consequently, the resulting scatter that contributes to the image is a function of the magnitude of the scattered radiation produced, its angular distribution and the density distribution of the object.
This motivates models in which a convolutional kernel is utilized to capture the spatial aspects of scatter,
but where the kernel somehow changes in response to the density distribution of the object,
e.g.,~\cite{sun_improved_2010}.

There exists a vast literature on scatter correction,
beginning in at least 1976~\cite{stonestrom_scatter_1976}
and presented in the context of many different scanners and applications,
from keV CT for medical imaging to MeV CT for nondestructive testing
(where keV and MeV refer to the nominal energy of the X-ray photons).
Here, we aim to discuss the main trends in computational approaches to scatter correction,
see the pair of reviews~\cite{ruehrnschopf_general_2011,ruehrnschopf_general_2011a}
for more comprehensive coverage.

Most scatter correction techniques rely on scatter estimation,
and can therefore be classified by 
what information they estimate the scatter from,
e.g.,
the total transmission radiograph,
a radiograph without scatter
(which we refer to as a direct radiograph),
or the reconstructed density~\cite{ruehrnschopf_general_2011}.
Scatter estimation from a reconstructed density volume can be highly accurate,
because, in principle, the density fully determines the scatter.
In practice, it is slow because it
necessitates alternating between
approximate reconstruction of the 3D density
and scatter estimation and removal.
Scatter estimation from the total transmission
(i.e., the radiographic measurements corrupted by scatter)
has the advantage of providing single-shot
descattering: no iteration is required;
however, these methods have limited accuracy
because some approximation is involved when predicting scatter
without access to the underlying density.
Scatter estimation from a radiograph without scatter
strikes a balance between accuracy and speed,
requiring only iteration in the radiograph domain
(the scatter is repeatedly estimated and removed based on iteratively estimated direct radiographs)
thereby avoiding the need to perform multiple
high-dimensional density reconstructions.
Which scatter estimation approach is best depends on the application:
the amount and dominant mechanism of scatter,
the requirements on descattering and reconstruction time,
and the desired accuracy of the reconstructions.

In this paper,
we propose a locally learned model of scatter,
wherein the free parameters of a given model are fit on training data
in a local neighborhood around the 2D radiograph to be descattered.
We emphasize that the proposed model is local in the sense that
the model changes as a function of the input---as distinct from spatially local models,
where the model changes as a function of location in the radiograph.

We use the proposed local scatter model as a component of an algorithm to reconstruct the material density field
of an object (with known material composition) from radiographs obtained with a polyenergetic X-ray beam.
Our focus on \emph{quantitative} material density reconstruction 
is somewhat different than in typical medical applications,
where the material composition of the object being imaged
(and therefore the physical properties that would allow mapping linear attenuation to density)
is unknown and therefore typically only the linear attenuation map is reconstructed
(with some exceptions, e.g., \cite{mason_polyquant_2017}).
The use of a polyenergetic beam introduces another complication to our problem
as compared to medical imaging, where often
(although not always, e.g., \cite{elbakri_statistical_2002})
a monoenergetic source is assumed.
In our setting, we assume known materials,
which gives us the opportunity to reconstruct density without 
facing an underdetermined problem.
Future work will examine objects consisting of multiple material compositions.

The rest of the manuscript is organized as follows.
The remainder of this section discusses related work.
In Section~\ref{sec:mono_and_poly}, we discuss the descattering problem,
including the forward model for monoenergetic and polyenergetic tomography.
In Section~\ref{sec:algo},
we discuss local modeling in general, 
then describe local models for scatter estimation and 
our descattering algorithm.
In Section~\ref{sec:experiments},
we present experiments validating our proposed local modeling approach.
We provide additional discussion in Section~\ref{sec:discussion}
and conclude in Section~\ref{sec:conclusion}.

\section{Monoenergetic and polyenergetic tomography with scatter}
\label{sec:mono_and_poly}
The overall goal of this work is to recover
a discretized, space-varying density
$\dens \in \mathbb{R}^{N_1\times N_2 \times N_3}$,
from a set of $V$ X-ray projection measurements (radiographs),
$\tot_1$, $\tot_2$, \dots, $\tot_V \in \mathbb{R}^{M_1\times M_2}$,
where $v$ indexes the viewing direction.
(If the object being imaged is axially symmetric,
as is the case in some important security applications,
a single view suffices for reconstruction,
since, ignoring noise, all views are identical).
In this section, 
we describe the forward model that relates $\dens$
to $\{\tot_v\}_{v=1}^{V}$;
much of our presentation is adapted from the standard textbook \cite{kak_principles_2001_nocity}.

\begin{figure}[htbp]
    \centering
    \begin{adjustbox}{scale=.8}
    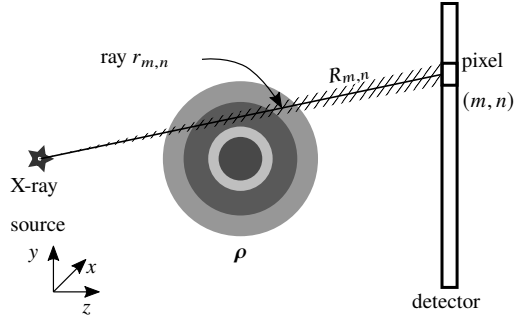
    \end{adjustbox}
    \caption{Sketch of the tomography setup.
    In reality, the object is 3D and the detector is a plane.
    $\dens \in \mathbb{R}^{N_1\times N_2 \times N_3}$ denotes the density field of the object being imaged,
    $R_{m,n}$ is the set of rays entering detector pixel $(m,n)$,
    and $r_{m,n}$ is the central ray of $R_{m,n}$.}
    \label{fig:setup}
\end{figure}

In a generic radiographic setup
(see Figure~\ref{fig:setup} for a sketch),
there is an X-ray source,
one or more objects being imaged,
and a detector.
Every ray between the source and detector, $r$,
intersects a certain amount of each object.
This is called the areal density of the object along $r$,
\begin{equation} \label{eq:areal}
    \areal_i(r) = \int_{-\infty}^\infty \rho_i(r_x(t), r_y(t), r_z(t)) dt,
\end{equation}
where $\areal_i(r)$ has units g/cm$^2$,
$r_x(t), r_y(t), r_z(t)$ are a parameterization of the ray $r$,
and
$i$ is the object index---%
each different material being imaged corresponds to a different $i$.
We note a slight abuse of notation here:
a continuous $\rho$ in \eqref{eq:areal}
versus a discrete $\dens$ mentioned before;
we will put aside the details of discretization here
as they are an implementation detail of the reconstruction software discussed in 
 Section~\ref{sec:reconstruction}. 
 
\subsection{Monoenergetic case} 
In the absence of scatter and with a monoenergetic source,
the number density of photons reaching the detector along ray $r$
is
\begin{equation}
\label{eq:beer}
    \photons(r) = \photons_{\text{in}}(r)
    \exp \left(-\sum_i \MAC_i \areal_i(r)\right),
\end{equation}
where $\photons(r)$ has units 1/cm$^2$,
$\photons_{\text{in}}(r)$ is the number density of the incident beam
along ray $r$,
and
$\MAC_i \in \mathbb{R}$ is the mass attenuation coefficient of the $i$th material,
with units cm$^2$/g.
In our application,
we assume known materials,
therefore the mass attenuation coefficients can be looked up,
e.g., in the XCOM database~\cite{berger_xcom_2010}.

In reality, 
what we measure is not the number density of photons at infinitesimal points,
as given by \eqref{eq:beer},
but a discrete, finite image 
which represents the number of photons reaching each of a finite grid of detectors.
We refer to this scatter-free measurement data 
as the direct radiograph, $\dir \in \mathbb{R}^{M_1 \times M_2}$.
It is given by
\begin{equation} \label{eq:direct-mono}
    \dir[m, n] 
    = \int_{R_{m,n}} \photons(r) dr
    \approx C \photons(r_{m,n})
\end{equation}
where $\dir[m,n]$ is unitless,
$R_{m,n}$ is the set of rays entering detector pixel $(m,n)$,
$r_{m,n}$ is the ray through the center of pixel $(m,n)$,
and
$C$ is a constant related to the area of a detector pixel and angle of incidence;
the approximation holds
when the detector elements are sufficiently small and regular in size
(think pixels).
Real measurement systems also have a per-pixel gain,
meaning that we actually measure $\gain[m,n]\dir[m,n]$.
We ignore this gain factor as it is easily handled via calibration.

\subsection{Polyenergetic case}
In the polyenergetic case, we have
\begin{equation}
\label{eq:beer-poly}
    \photons(r) = \int_0^{\infty}
    \photons_{\text{in}}(r, E)
    \exp \left(-\sum_i \MAC_i(E) \areal_i(r)\right)
    dE,
\end{equation}
where 
$\photons_{\text{in}}(r, E)$ is the energy profile of the number density of the incident beam
and
$\MAC_i(E)$ gives the mass attenuation coefficient
of the $i$th material at energy $E$.
Using the same numerical integration approximation as in the monoenergetic case
and an additional quadrature for energy
gives
\begin{equation} \label{eq:direct-poly}
    \dir[m, n] 
    \approx \sum_e C \photons(r_{m,n}, E_e),
\end{equation}
where $e$ indexes discrete energy bins,
$C$ is a constant depending on the pixel size and energy bin width,
and
    \begin{equation} \label{eq:q-poly}
        \photons(r, E) = 
        \photons_{\text{in}}(r, E)
        \exp \left(-\sum_i \MAC_i(E) \areal_i(r)\right).
    \end{equation}

\begin{figure*}
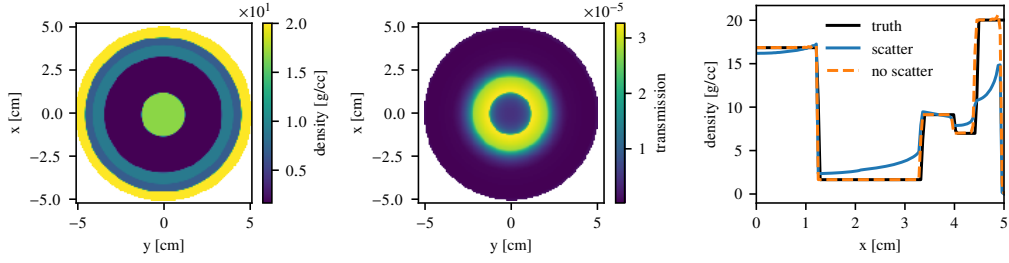

    \centering
    \begin{adjustbox}{scale=.75}
    \input{figures/mono_rho_im.pgf}%
    \end{adjustbox}%
    \begin{adjustbox}{scale=.75}    
    \input{figures/mono_total_im.pgf}%
    \end{adjustbox}%
    \begin{adjustbox}{scale=.75}%
    \input{figures/mono_recon_profiles.pgf}%
    \end{adjustbox}     
    \caption{Illustration of how the presence of scatter negatively impacts quantitative density reconstruction (monoenergetic dataset).
    Left: Cross section showing the density of a spherically symmetric, 3D object.
    Right: Radiographic measurement of the object on the left, which includes scatter.
    Our goal is to reconstruct the left image from the center one.
    Bottom: 1D profiles of the true density
    and reconstruction from radiographs with and without scatter.
    Without scatter, reconstruction is nearly perfect
    (orange, dotted line overlaps black);
    with scatter, reconstruction involves density errors on the order of 1 g/cm$^3$.}
    \label{fig:data_example}
\end{figure*}

In some cases,
it is sufficient to stop at
the model for the direct radiograph given by
\eqref{eq:direct-mono} or \eqref{eq:direct-poly}.
However, in many cases,
there is a nontrivial amount of scatter,
i.e.,
photons that reach the detector
without following a straight path through the object
(Figure~\ref{fig:data_example} illustrates the effect this scatter on reconstruction).
Scatter arises from several different physical processes.
A photon may
    undergo exactly one coherent scattering interaction with the object;
    undergo exactly one Compton scattering interaction with the object;
    interact with and scatter from parts of the measurement apparatus, e.g., the detector itself~\cite{bhatia_convolution_2017} or the room;
    or
    undergo any other interaction(s) not included above, e.g.,
    multiple scatters or pair production.
These processes are well understood, 
but they are complicated enough that they need to be modeled
at the individual photon level,
a computationally expensive process
and one that is difficult to invert.
For our purposes,
we consider the measured radiograph
$\tot \in \mathbb{R}^{M_1 \times M_2}$
as a sum of direct photons, $\dir$
as described in \eqref{eq:direct-mono} or \eqref{eq:direct-poly},
and scattered photons, $\sca \in \mathbb{R}^{M_1 \times M_2}$,
$\tot = \dir + \sca$.
We will also refer to $\tot$ as the total radiograph,
since it contains both the contribution of the scattered and direct photons.
Finding a way to efficiently and accurately model $\sca$
is the main focus of this work.

\section{Proposed approach---local modeling for descattering} \label{sec:algo}
As stated earlier, 
the goal of this work is to recover a 3D object,
$\dens \in \mathbb{R}^{N_1\times N_2 \times N_3}$,
from its 2D tomographic measurements (corrupted by scatter),
$\tot_1$, $\tot_2$, \dots, $\tot_V \in \mathbb{R}^{M_1\times M_2}$.
Our proposed approach consists of two steps:
descattering radiographs and tomographic reconstruction;
we now discuss the first of these.
We discuss related work,
introduce our proposed local modeling approach,
 describe the specific local models we evaluate,
and finish with a description of our descattering algorithm.

\subsection{Related work on scatter estimation}
Most methods that estimate scatter from a direct radiograph use some form 
of the model
\begin{equation} \label{eq:general-model}
\sca[m,n] = \sum_{m',n'} f(\dir[m',n']) \vec{H}_\dens[m', n', m, n],
\end{equation}
where $\sca[m,n]$ and $\dir[m,n]$ are the scatter (only) and direct (without scatter) radiographs at position $m,n$
and
$f$ is a scalar function.
The term $f(\dir[m,n])$, sometimes called the
\emph{scatter potential} or \emph{scatter amplitude factor},
represents how much scatter comes off of a given pixel.
The term $\vec{H}_\dens[m', n', m, n]$,
called the \emph{scatter kernel},
represents how much radiation scatters from position $(m',n')$
to position $(m, n)$.
In the most general case,
it is dependent on the 3D density field of the object being imaged, $\dens$.

The model \eqref{eq:general-model} is hardly useful as it is because it is so general.
The challenge is simplifying it to a point where it can be parameterized and the parameters estimated.
One common approach is to 
fix the scatter amplitude factor to the parametric form
$f(x) = x^\alpha | \ln x |^\beta$,
which can be motivated by a simple physical model~\cite{ohnesorge_efficient_1999}.
Another is to constrain the scatter kernel to be shift invariant, $\vec{H}_\dens[m', n', m, n] = \vec{H}_\dens[m'-m, n'-n]$,
or a spatially-varying weighted sum of such kernels,
where the weights may depend on (estimated) object thickness
or distance to the object boundary~\cite{sun_improved_2010}.
There also exists recent work that uses a
convolutional neural network to map from a direct radiograph to the corresponding scatter radiograph~\cite{maier_deep_2018},
which we can view as an overparameterized and nonlinear version of \eqref{eq:general-model}.

No matter the form of the scatter model,
its parameters must be determined in order to perform scatter estimation.
One approach is to fit the parameters on simple simulations,
e.g., of a pencil beam through a slab~\cite{sun_improved_2010, tisseur_evaluation_2018}.
Another is to approach fitting as an optimization problem on complex reference simulation(s)~\cite{maier_deep_2018}.
The benefit of the former approach is that it may generalize well
and fits well with physically interpretable models.
The benefit of the latter is that it may provide more accurate fits,
because it explicitly minimizes fit error.

\subsection{Local modeling formulation}
Given a training set of 2D direct radiographs
and the corresponding 2D scatter radiographs,
$\{(\dir_t, \sca_t )\}_{t=1}^{T}$,
our goal is to learn a mapping,
$\model: \mathbb{R}^{M_1 \times M_2} \to \mathbb{R}^{M_1 \times M_2}$,
from direct to scatter.
(We leave aside the question of whether such a mapping exists,
i.e., whether two objects can produce the same direct radiograph but different
scatter radiographs.)
These training radiographs can be formed by generating synthetic objects, $\rho_t$,
and simulating their radiographs via particle transport software,
e.g., MCNP~\cite{werner_mcnp6.2_2018} or Geant4~\cite{agostinelli_geant4_2003}.%
See Section~\ref{sec:experiments:data} for details of the training sets we used in our experiments.

Our proposed approach is to represent $\model$
via a large number of simple models,
i.e., a collection of $\model_{\dir}$'s,
each acting in a different region of $\mathbb{R}^{M_1 \times M_2}$.
To do this,
we define the nearest neighbor function, $\mathbb{U}$,
to return the set of indices of the $G$ nearest training pairs to $\dir$ in the training set,
i.e.,
$\mathbb{U}(\dir) = \{t_1, t_2, \dots t_G\},$
such that $\| \dir - \dir_s \|_F^2 \le \| \dir -  \dir_t \|_F^2$ for $s \in \mathbb{U}(\dir)$
and $t \not\in \mathbb{U}(\dir)$,
where $\| \cdot \|_F^2$ denotes the squared Frobenius norm 
(sum of the squared elements).
(Other metrics may be used to determine these neighborhoods, 
e.g., the Wasserstein metric, if they are more suitable to the particular application)
Finally, 
we select a class, $\mathcal{H}$, of models for the local $\model_{\dir}$'s;
see Section~\ref{sec:models} for specific examples.

Our local model $\model$
is then given by
\begin{equation} \label{eq:local_fit}
     \model(\dir) 
     = \model_{\dir}(\dir),
     \quad \text{where} \quad
     \model_{\dir} 
     = \argmin_{\mathcal{B} \in \mathcal{H}} \sum_{t \in \mathbb{U}(\dir)}
     \|\mathcal{B}(\dir_t)
     - \sca_t \|_F^2.
\end{equation}
Note the use of a subscript $\dir$:
$\model_\dir$ is a model (i.e., an image-to-image function) fit on the neighbors of $\dir$;
$\model_\dir(\dir)$ is that model evaluated at the point $\dir$,
which is an image.
Put simply:
to predict $\sca$ from $\dir$,
we fit a model on the nearest neighbors to $\dir$
in the training set
and apply that model to $\dir$.

What we have described is distinct from the standard approach to fitting a model on a training set,
which we refer to here as \emph{global} modeling.
Using our formulation,
global modeling corresponds to local modeling with
the number of neighbors, $G$, set to the entire training set size, $T$.
This simplifies to 
\begin{equation} \label{eq:global_fit}
     \model_{\dir} = \model 
     = \argmin_{\mathcal{B} \in \mathcal{H}} \sum_{t=1}^{T}
     \|\mathcal{B}(\dir_t)
     - \sca_t \|_F^2.
\end{equation}
To see the conceptual advantage of local over global models,
consider the case of linear regression:
global modeling fits a single hyperplane though all of the data,
while local modeling fits different hyperplanes in different regions of the space.
Thus, local modeling can accurately describe more complex relationships than global modeling,
with the drawback that it may lead to worse overfitting.

\textbf{Related work on local modeling.} Local modeling has a long history in a variety of fields.
For example,
in statistics, local regression (also called LOESS),
which involves fitting local models to form a smooth interpolator,
has been studied since at least the 1970s~\cite{cleveland_robust_1979}.
The novelty of our work is to propose and validate specific,
high-dimensional local models for use in the context of X-ray scatter estimation,
descattering,
and density reconstruction.

\subsection{Models}
\label{sec:models}
We now describe the four classes of models 
(i.e., choices of $\mathcal{H}$ in \eqref{eq:local_fit} and  \eqref{eq:global_fit}) used in our experiments.
The first two (single field and convolutional) are simple models that we propose;
the second two (parametric and multikernel) come from the scatter modeling literature~\cite{sun_improved_2010}.

\subsubsection{Single field model}
The single field model represents the scatter as being a single image,
$\model_\dir(\dir) = \hat{\sca}_\dir \in \mathbb{R}^{M_1 \times M_2}$.
Fitting the single field model requires solving the problem
\begin{equation} \label{eq:single_fit}
    \hat{\sca}_\dir = 
    \argmin_{\sca \in \mathbb{R}^{M_1 \times M_2}}
    \sum_{t \in \mathbb{X}}
     \| \sca
     - \sca_t \|^2_F,
\end{equation}
where $\mathbb{X}$ may represent either the entire training set (\emph{global fitting})
or $\mathbb{U}(\dir)$ (\emph{local fitting}).
The solution to \eqref{eq:single_fit} is easily obtained
by taking the pixelwise average of the $\{\sca_t\}_{t \in \mathbb{X}}$.
In the case where one neighbor is used,
the single field model is the classical nearest neighbor interpolator.
This model provides a simple baseline for comparison.

\subsubsection{Convolutional model}
The convolution model represents the scatter as a convolution,
which is a simple and computational efficient way of modeling the spatial distribution of scatter.
The form of the model is $\model(\dir) = \ker_\dir \ast f_{1,1}(\dir)$,
where
\begin{equation} \label{eq:nonlinear}
    f_{\alpha, \beta}(\dir)[m_1, m_2] = -\dir[m1, m2]^\alpha (\log \dir[m_1,m_2])^\beta.
\end{equation}
Here, $\ast$ denotes 
the 2D, linear convolution between zero-padded images,
where the result is cropped to match the support of $\dir$.
The size of the kernel
is the smallest such that each point in $\dir$
can contribute to each point in the result,
thus $\ker_\dir \in \mathbb{R}^{(2M_1-1) \times (2M_2-1)}$.
Each pixel in the kernel is arbitrary, thus, the number of parameters for the convolutional model is $(2M_1-1) \times (2M_2-1)$.
The nonlinear function $f$ is common in the descattering literature
and can be justified by a physics argument when $\alpha=\beta=1$~\cite{ohnesorge_efficient_1999}.
(To make sense of the negative sign, 
note that $\dir$ is expected to be normalized 
to lie between zero and one.)
The flexibility provided by $\alpha$ and $\beta$ is used in the next model.

The kernel is determined by solving the optimization problem
\begin{equation} \label{eq:kernel_fit}
    \ker_\dir = \argmin_\ker
    \sum_{t \in \mathbb{X}} \| \ker \ast   f_{1,1}(\dir_{t}) - \sca_{t} \|_F^2
\end{equation}
where
$\mathbb{X}$ is an appropriate set of training indices (local or global).
The problem \eqref{eq:kernel_fit} is a linear least squares problem
because the $f_{1,1}(\dir_{t})$'s are fixed.
For our experiments, 
we solved it using the conjugate gradient method~\cite{shewchuk_introduction_1994}.

\subsubsection{Parametric kernel model}
The parametric kernel model is a version of the convolution model
with a parametric kernel and nonlinearity.
It appears in the recent literature on descattering, e.g., \cite{sun_improved_2010}.
The form of the model is
$\model(\dir)
    = \ker_{A, B, \sigma_1, \sigma_2} \ast 
    f_{\alpha, \beta}(\dir),$
where $f$ is defined as in \eqref{eq:nonlinear}
and all of the parameters, $A$, $B$, $\sigma_1$, $\sigma_2$, $\alpha$, and $\beta$
depend on $\dir$.
The kernel $\ker$ is a sum of two centered Gaussian functions,
\begin{equation} \label{eq:two_gaussians}
    \ker_{A, B, \sigma_1, \sigma_2}[m, n] = 
    \frac{A}{\sigma_1 \sqrt{2 \pi}}
    e^{\left(-\frac{(m-m_0)^2 + (n-n_0)^2}{2\sigma_1^2}\right)} + \\
     \frac{B}{\sigma_2 \sqrt{2 \pi}}
    e^{\left(-\frac{(m-m_0)^2 + (n-n_0)^2}{2\sigma_2^2}\right)},
\end{equation}
where $m_0, n_0$ gives the central pixel of the radiograph.
Fitting the parametric kernel model involves the following nonconvex optimization with respect to the scalar parameters $A, B, \sigma_1, \sigma_2, \alpha$, and $\beta$,
\begin{equation}
\label{eq:parametric_fit}
    \argmin_{A, B, \sigma_1, \sigma_2, \alpha, \beta}
    \sum_{t \in \mathbb{X}}
     \| \model(\dir)
     - \sca_t \|^2_F.
\end{equation}
For our experiments, we used the LBFGS algorithm~\cite{liu_limited_1989} (using the PyTorch implementation),
which we found to be significantly faster than gradient descent.
We note that LBFGS is not guaranteed to find the global minimizer of $\eqref{eq:parametric_fit}$
because it is nonconvex.

\subsubsection{Multikernel model}
The multikernel model,
proposed in \cite{sun_improved_2010},
is a generalization of the parametric kernel model
wherein different parameters are used in different regions of the image. 
To fit $K$ kernels, $K-1$ thresholds, $\tau_1, \tau_2, \dots, \tau_{K-1}$ are established
and a partition function is defined according to
\begin{equation}
    \vec{r}_\dir[m,n] = 
    \begin{cases}
    1 & 0 \le -\log(\dir[m,n]) \le \tau_1\\
    2 & \tau_1 < -\log(\dir[m,n]) \le \tau_2\\
    \vdots & \vdots \\
        K & \tau_{K-1} < -\log(\dir[m,n]).
    \end{cases}
\end{equation}
The model is then given by
\begin{equation}
    \model(\dir)
    = \sum_{k=1}^K -\ker_{A_k, B_k, \sigma_{1,k}, \sigma_{2,k}}
    \ast 
    (\vec{1}_{\vec{r}_\dir=k} \odot f_{\alpha_k, \beta_k}(\dir)),
\end{equation}
where each kernel is given by \eqref{eq:two_gaussians},
the indicator function is given by
\begin{equation}
    \vec{1}_{\vec{r}=k}[m ,n]=
    \begin{cases}
    0, & \vec{r}[m, n] \ne k; \\
    1, & \vec{r}[m, n] = k,
    \end{cases}
\end{equation}
and $\odot$ indicates elementwise multiplication.
Put simply, the multikernel model uses a combination of parametric kernel models,
each acting in a different region of the input (specified by the indicator).
Fitting the multikernel model, like the parametric kernel,
involves solving a nonconvex optimization problem
over all the scalar parameters,
for which we again used the LBFGS algorithm~\cite{liu_limited_1989}.

\subsection{Descattering algorithm}
\label{sec:descattering}
We have so far discussed scatter estimation from a known direct radiograph,
but our ultimate goal is descattering,
estimating a direct radiograph from a total radiograph containing scatter.
With the scatter models from Section~\ref{sec:models} in hand, 
we are ready to descatter by solving
     $\tot = \model(\dir) + \dir$
for $\dir$.
(One could readily incorporate regularization on $\dir$ and terms modeling noise or other errors in this model;
we leave these for future work.)
Inverting this equation may be challenging
because it involves additional underlying optimization problems such as
\eqref{eq:single_fit}, \eqref{eq:kernel_fit}, or \eqref{eq:parametric_fit}
that can themselves be challenging and nonconvex.
Here, we use an iterative fixed-point method.
Fixed-point algorithms are commonly used for descattering, e.g., see \cite{sun_improved_2010}.
The wrinkle here is that, in the local learning case,
the scatter model parameters themselves
change as a function of the current estimate of $\dir$.
The pseudocode for our descattering algorithm is given in Algorithm~\ref{algo:fixed_point}.
\begin{algorithm}
\begin{algorithmic}[1]
\Function{descatter}{$\tot$, $\{(\dir_t, \sca_t)\}_{t=1}^T$}
    \State initialize: $\dir \gets \tot$,
    \For{fixed number of iterations}
    \State fit a model, $\model_\dir$
        \Comment{e.g., via \eqref{eq:single_fit}, \eqref{eq:kernel_fit},
        or \eqref{eq:parametric_fit}}
        \State $\dir \gets \tot - \model_\dir(\dir)$
        \State $\dir[\dir < 0] \gets 0$
            \Comment{enforces nonnegativity of $\dir$}
    \EndFor
    \State \textbf{return} $\dir$
\EndFunction
\end{algorithmic}
\caption{Fixed point descattering.}
\label{algo:fixed_point}
\end{algorithm}

\subsection{Monoenergetic and polyenergetic density reconstruction} \label{sec:reconstruction}
After descattering, we have 
an estimate of the set of direct radiographs, $\{\dir_v\}_{v=1}^V$,
wherein each $\dir_v$ is estimated independently.
The next tasks is to reconstruct the density field of the object, $\dens$
from this set,
which amounts to inverting \eqref{eq:direct-mono} or \eqref{eq:direct-poly}
(and, consequently, \eqref{eq:beer} or \eqref{eq:beer-poly} and \eqref{eq:areal})
with respect to the underlying density.
In general, if the inversion is ill-posed (e.g., limited view imaging),
additional regularization can be used during reconstruction~\cite{mccann_biomedical_2019,ravishankar_image_2020}.
Here, we consider the case when enough views are acquired to adequately reconstruct the 3D object.
In particular, because the objects we are imaging,
which are of interest in important nuclear security applications, are spherically symmetric,
a single view suffices for reconstruction~\cite{dasch_one_1992}.
This inversion is a well-studied problem and not the main focus of this work;
we therefore leave the details of our approach to Sections~\ref{sec:recon_mono} and \ref{sec:recon_poly}.

\section{Experiments and results} \label{sec:experiments}
To evaluate our algorithms,
we performed descattering and reconstruction experiments with simulated data.
These experiments elucidate the descattering performance of the methods,
their ability to fit to known scatter,
and the effect the number of local neighbors used.
In this section, we describe implementation details for each comparison method,
our datasets,
and our experiments and results.

\subsection{Preliminaries}
\label{sec:prelim}
\textbf{Implementation details.}
In all experiments, scatter models were fit as described
in Section~\ref{sec:models},
and all code was written in Python.
The area outside the object, i.e., beyond a 5 cm radius,
was excluded in all cost functionals
because, in practice,
there would be no difficulty locating the outer edge of the object.
Full-resolution images were $257 \times 257$,
but scatter fitting was performed on $4\times$ downsampled versions
(to improve fitting speed and because the scatter is spatially smooth;
this is common in the literature, e.g., ~\cite{sun_improved_2010})
and scatter estimates were upsampled to the original size using bilinear interpolation.
Data was rescaled to simplify the tuning of fitting routines.
Specifically:
each training direct radiograph was divided by the Frobenius norm of the entire direct training set (square root of the sum of the squares of all pixels).
Each training scatter radiograph was divided by the Frobenius norm of the entire scatter training set.
At testing time, 
model inputs were divided by the previously-computed training direct norm and outputs were multiplied by the previously computed training scatter norm.
For CG (for the convolutional model),
the number of iterations was 40.
For both the parametric kernel and multikernel models,
the number of LBFGS iterations was 20 with step size 1.0
and back-tracking line search was enabled;
the initial parameters were $A=1.0, B=1.0, \sigma_1=4, \sigma_2=64, \alpha=1.  0, \beta=0$.
For the multikernel model, we used $K=3$.
While we did not make a careful study of the runtimes of the proposed methods,
fitting global models was slower than fitting local ones
(minutes versus seconds on a consumer-level laptop)
and evaluating the models once fit took under a second for all models.
Code for our method is available upon request.

Inverse Abel transforms were performed with the three-point algorithm 
of \cite{dasch_one_1992} from the PyAbel Python package.
Polyenergetic reconstructions were slower than monoenergetic ones
due to the lookup table,
but both took on the order of seconds on a consumer-level laptop.

\textbf{Box and whisker plots.}
Throughout this section,
we summarize results using box and whisker plots,
e.g., see Figure~\ref{fig:descattering}.
Unless otherwise noted,
each box summarizes the ten testing images.
Whiskers indicate minimum and maximum values 
and boxes indicate first and third quartile.
The orange line in the center of each box indicates the median value.

\subsection{Datasets}
\label{sec:experiments:data}
Our experiments use two simulation datasets with properties motivated
by our nuclear security application:
large, dense, spherical objects.
We chose to use objects comprising a single material (plus a known collimator)
to simplify reconstruction and focus on the effects of descattering.

We generated synthetic radiographs using Monte Carlo particle transport simulation software
(MCNP6~\cite{werner_mcnp6.2_2018}) that consisted of five concentric shells.
The outermost shell had a fixed radius of 5 cm and the other four shell radii were randomly varied between 0.25 and 2.5 cm.
The material was set to uranium for the entire object,
but the density of each shell was randomly selected from a set of five densities, 
and the mass of the object was conserved through a multiplicative constant.
The object center was placed 133 cm from the source and the detector was placed at 525 cm from the source.
A tantalum collimator (3.5 inches thick, 6.4 inch diameter, 1.4 inch opening, and 2.55 degree half-angle) was placed at 98 cm from the source. 
While the collimator design was not modified for each object, 
it served to reduce the dynamic range of the image, in general. 
We simulated radiographs of these objects with both a 
monoenergetic and polyenergetic X-ray source.
The monoenergetic single-material dataset was generated using a 1.5 MeV X-ray source and
the polyenergetic single-material dataset was generated using a bremsstrahlung with a 19.4 MeV endpoint energy. 
We generated synthetic radiographs using the radiography tally in MCNP, which generates and image produced by X-rays.  This tally is a quasideterministic calculation that maps each particle to each pixel and the final image does not contain stochastic noise.  Radiographs of direct and scattered radiation were generated separately.  The direct radiation calculations utilized 50000 particles while the scattered radiation calculations utilized 1e6 particles.  Additionally, the scattered radiation calculations included the PDS card in MCNP, which estimates all possible reaction contributions for each collision and significantly reduces the appearance of fliers due to rare coherent scatter events.  Degradation of the image due to source blur and detector effects such as scintillator efficiency and blur were not modeled in this study.

For the monoenergetic dataset, we generated 99 pairs of direct and scatter radiographs.
For the polyenergetic dataset, we generated 100 pairs.
These were the largest datasets we could simulate in a reasonable amount of (order of days).
In both cases, we partition the datasets into a testing set consisting of ten pairs
(used for evaluation of algorithms)
and a training set consisting of the remaining pairs
(used for fitting models).
These partitions were chosen to balance the need for computing statistics on the testing set with the desire to have as large a training set as possible.

\subsection{Descattering and density reconstruction}
\label{sec:exp:descatter}

\begin{figure}[htbp]
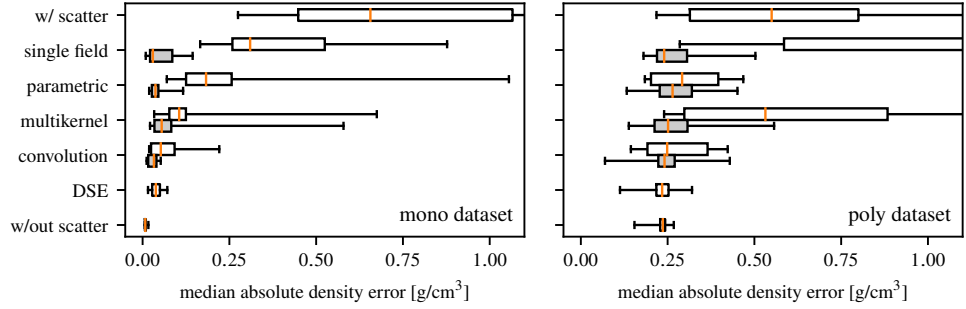

    \centering
    \begin{adjustbox}{clip,trim=0.275cm 0.25cm 0cm 0cm,scale=0.9}%
    \input{figures/density_error_descatter_mono.pgf}%
    \end{adjustbox}%
     \begin{adjustbox}{clip,trim=1.8cm 0.25cm 0cm 0cm,scale=0.9}%
    \input{figures/density_error_descatter_Brem.pgf}%
    \end{adjustbox}%
    \caption{Error in density reconstruction after descattering for several scatter estimation methods,
    local and global versions.
    In each stacked pair of boxes, the upper, white box is the global version and
    the lower, grey box is the local version.
    For the sake of comparison,
     the  top- and bottom-most boxes  give  the  error  when scatter is not removed
     and when it is removed perfectly, respectively.
    Local methods always outperformed their global counterparts.
    }
    \label{fig:descattering}
\end{figure}

\begin{figure*}[htbp]
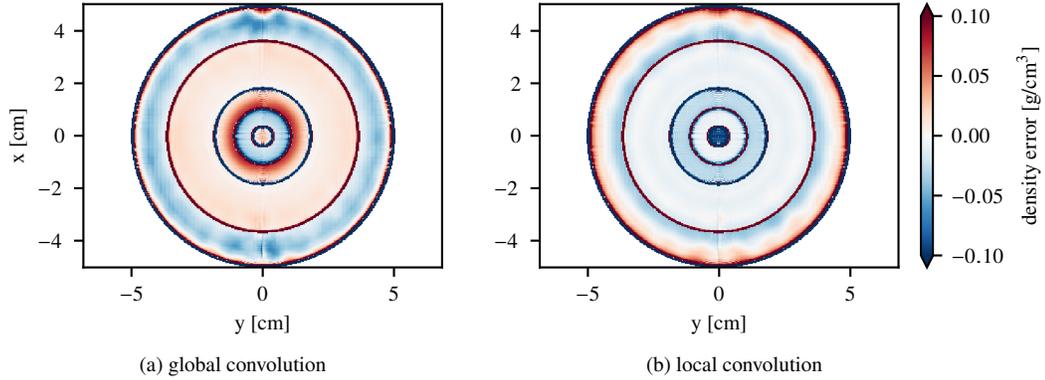

\centering
\begin{subfigure}{.5\linewidth} 
\begin{adjustbox}{clip,trim=0cm .3cm 2cm .2cm}
\input{error_im_mono_convolution_global.pgf}%
\end{adjustbox}%
\caption{global convolution}           
\end{subfigure}%
\begin{subfigure}{.5\linewidth}
\begin{adjustbox}{clip,trim=.6cm .3cm 0.15cm .2cm}%
\input{error_im_mono_convolution_local.pgf}%
\end{adjustbox}%
\caption{local convolution}  
\end{subfigure}
\caption{Error magnitude in density reconstruction after descattering for one representative object
    in the monoenergetic dataset.}
    \label{fig:descatter_error_im}
\end{figure*}

Our main experiment evaluated the ability of 
each of the models 
to enable accurate descattering
and density reconstruction.
The experimental setup mimicked our actual application.
We fit scatter models on a training set of $(\dir_t, \sca_t)$ pairs
and then
descattered a set of new total radiograph using 
the fit models and Algorithm~\ref{algo:fixed_point}.
Finally, we reconstructed the underlying density based on the estimated
direct radiograph.
For local models, we fixed the number of neighbors to three
(we explore the effect of the number of neighbors in Section~\ref{sec:neighbors}).

We quantified the reconstruction errors in terms of
the median absolute density error (MADE) on the central slice of the
3D reconstruction,
\begin{equation} \label{eq:MADE}
\median_{m,n}  \left(
\abs{\hat{\dens}_t[m,n,0] - \dens_t[m,n,0]}
\right),
\end{equation}
in  g/cm$^3$,
where $\hat{\dens}$ is the reconstructed density,
$\dens_t$ is the ground truth,
and
we let $m$ and $n$ range over the area
where $\dens_t[m,n,0]$, is nonzero.
(Recall that the objects we are imaging are 3D,
hence the three indices for $\dens$.)
We use the central slice it because matches one way we typically view the reconstructions, i.e., as a slice; quantifying error on 1D profiles or the whole volume would also be reasonable.
We used the median rather than the mean because 
the mean is dominated by errors at the edge of the simulation
and at the interface between materials.

\textbf{Comparison methods.}
While our main focus is the comparison between global fitting and the proposed local fitting,
we note that the global versions of the parametric model 
and multikernel model represent baseline comparison methods from the recent literature~\cite{sun_improved_2010}.
As as an additional comparison,
we implemented the deep scatter estimation (DSE) network~\cite{maier_deep_2018},
which is a CNN-based scatter estimation method
(trained globally and incorporated into the fixed point algorithm 
in the same manner as the other scatter models).

\begin{figure*}[!htb]
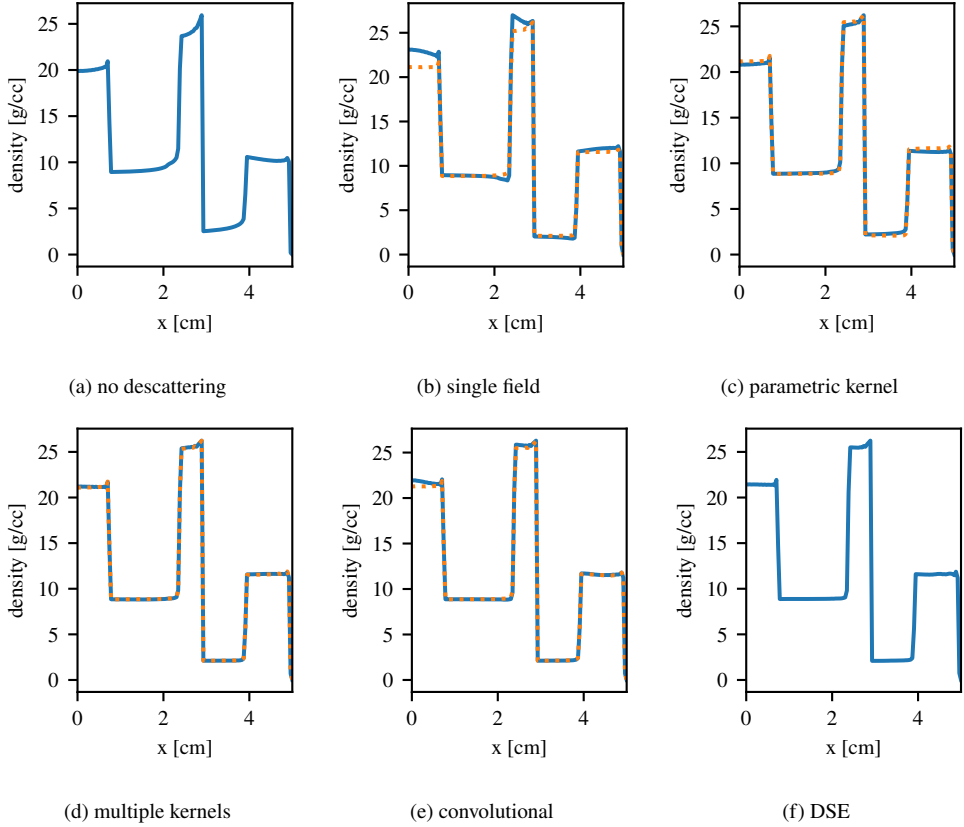

\captionsetup[subfigure]{justification=centering}
 \begin{subfigure}{.33\linewidth}
    \begin{adjustbox}{clip,trim=0.0cm 0cm 0cm 0cm}
    \input{figures/recon_profile_mono_no_descattering.pgf}%
    \end{adjustbox}
    \caption{no descattering}
    \end{subfigure}%
    \begin{subfigure}{.33\linewidth}
    \begin{adjustbox}{clip,trim=0.0cm 0cm 0cm 0cm}
    \input{figures/recon_profile_mono_single_field.pgf}%
    \end{adjustbox}
    \caption{single field}
    \end{subfigure}%
    \begin{subfigure}{.33\linewidth}
    \begin{adjustbox}{clip,trim=0.0cm 0cm 0cm 0cm}    
    \input{figures/recon_profile_mono_parametric_kernel.pgf}%
    \end{adjustbox}
    \caption{parametric kernel}
    \end{subfigure}\\
    \begin{subfigure}{.33\linewidth}
    \begin{adjustbox}{clip,trim=0.0cm 0cm 0cm 0cm}    
    \input{figures/recon_profile_mono_multikernel.pgf}%
    \end{adjustbox}
    \caption{multiple kernels}
    \end{subfigure}%
    \strut\hfill
    \begin{subfigure}{.33\linewidth}
    \begin{adjustbox}{clip,trim=0.0cm 0cm 0cm 0cm}    
    \input{figures/recon_profile_mono_convolution.pgf}%
    \end{adjustbox}
    \caption{convolutional}
    \end{subfigure}\hfill
    \begin{subfigure}{.33\linewidth}
    \begin{adjustbox}{clip,trim=0.0cm 0cm 0cm 0cm}    
    \input{figures/recon_profile_mono_DSE.pgf}%
    \end{adjustbox}
    \caption{DSE}
    \end{subfigure}\hfill\strut
    
    \caption{Example reconstruction profiles (monoenergetic dataset).
    Solid: global model;
    dotted: local model.
    See Figure~\ref{fig:descatter_error_profiles}
    for error plots where differences are more apparent.
    }
    \label{fig:recon_profiles}
\end{figure*}

\begin{figure*}[!htb]
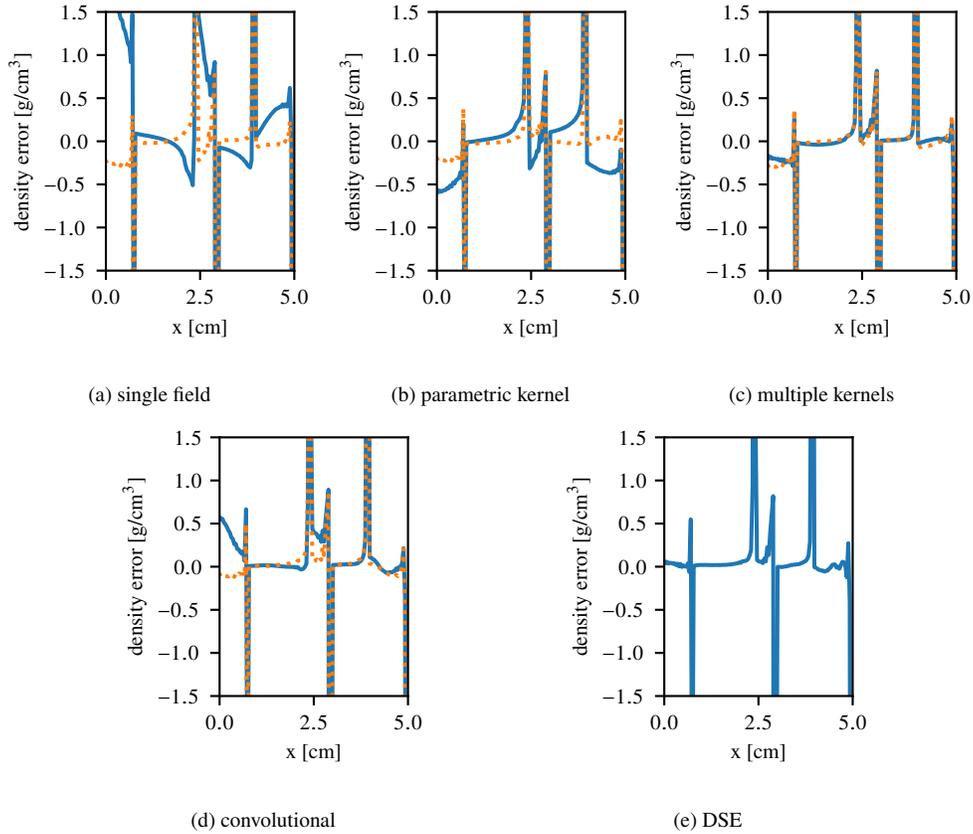

\captionsetup[subfigure]{justification=centering}
    \begin{subfigure}{.33\linewidth}
    \begin{adjustbox}{clip,trim=0.0cm 0cm 0cm 0cm}
    \input{figures/error_profile_mono_single_field.pgf}%
    \end{adjustbox}
    \caption{single field}
    \end{subfigure}%
    \begin{subfigure}{.33\linewidth}
    \begin{adjustbox}{clip,trim=0.0cm 0cm 0cm 0cm}    
    \input{figures/error_profile_mono_parametric_kernel.pgf}%
    \end{adjustbox}
    \caption{parametric kernel}
    \end{subfigure}%
    \begin{subfigure}{.33\linewidth}
    \begin{adjustbox}{clip,trim=0.0cm 0cm 0cm 0cm}    
    \input{figures/error_profile_mono_multikernel.pgf}%
    \end{adjustbox}
    \caption{multiple kernels}
    \end{subfigure}\\  
    \strut\hfill
    \begin{subfigure}{.33\linewidth}
    \begin{adjustbox}{clip,trim=0.0cm 0cm 0cm 0cm}    
    \input{figures/error_profile_mono_convolution.pgf}%
    \end{adjustbox}
    \caption{convolutional}
    \end{subfigure}\hfill
    \begin{subfigure}{.33\linewidth}
    \begin{adjustbox}{clip,trim=0.0cm 0cm 0cm 0cm}    
    \input{figures/error_profile_mono_DSE.pgf}%
    \end{adjustbox}
    \caption{DSE}
    \end{subfigure}\hfill\strut
    
    \caption{Example reconstruction error profiles (monoenergetic dataset).
    Solid: global model;
    dotted: local model.
    Descattering improves density reconstruction across the entire profile
    and
    local models generally outperform their global counterparts.
    }
    \label{fig:descatter_error_profiles}
\end{figure*}

\textbf{Results.}
The results for the descattering experiment are given in 
Figure~\ref{fig:descattering}.
The results 
show that local models uniformly outperform global ones,
which justifies the use of the proposed local fitting approach.
For the monoenergetic dataset, the best descattering
was provided by the local convolutional model (maximum MADE=0.040 g/cm$^3$),
followed by the local parametric model (0.045 g/cm$^3$)
and the DSE (0.050 g/cm$^3$).
None of the methods matched the performance
of the reconstruction without scatter (maximum MADE=.009 g/cm$^3$).
For the polyenergetic dataset,
the best descattering
was provided by the DSE (maximum MADE=0.253 g/cm$^3$),
followed by the local convolutional model (0.270 g/cm$^3$).
Here, the error when reconstructing from the direct signal (0.243 g/cm$^3$) is higher than
for the monoenergetic case:
this is to be expected because the particle transport simulation used to generate the data
and the forward model used to reconstruct it are distinct.
Slight differences
in the physical constants
(beam spectrum and mass attenuation coefficients)
between the model and simulation
cause these reconstruction errors.
We believe that this mismatch is a strength of the experiments,
because it reflects (at least partly) the uncertainties that exist when reconstructing from experimental data.
We show
error images for the convolutional method in Figure~\ref{fig:descatter_error_im}.
This shows that errors are highest at material interfaces,
and that there exist both density under- and overshoots.
We provide reconstruction and error profiles in Figures~\ref{fig:recon_profiles} and \ref{fig:descatter_error_profiles}. 

\begin{figure}[!htb]
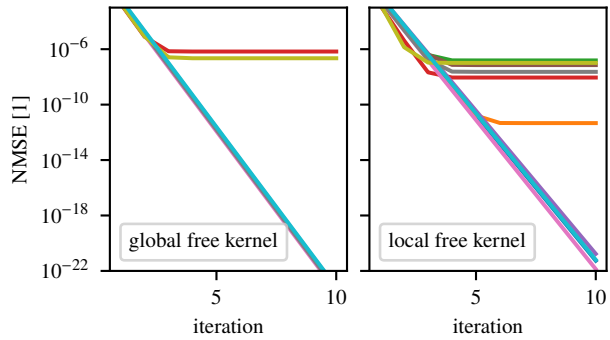

\captionsetup[subfigure]{justification=centering}
    \centering
    \begin{adjustbox}{clip,trim=.25cm .40cm .25cm .25cm}
    \input{figures/mono_convolution_global.pgf}
    \end{adjustbox}%
    \begin{adjustbox}{clip,trim=1.5cm .40cm .25cm .25cm}
    \input{figures/mono_convolution_local_2.pgf}
    \end{adjustbox}
    \caption{Convergence of the fixed point algorithm 
    using the convolutional model on the monoenergetic dataset.
    Each line represents one image in the test set,
    all ten are not visible due to overlaps.
    }
    \label{fig:convergence}
\end{figure}

\textbf{Convergence.}
We quantified the convergence of the fixed point algorithm
by computing the normalized mean squared error (NMSE) at each iteration,
$    \|\dir+\model_\dir(\dir) - \tot\|_2^2 / \|\tot\|_2^2.
    $
    (Note that this is not identically zero
    because the $\dir$s on the left and right hand sides of Algorithm~\ref{algo:fixed_point} line 5 represent two successive iterations and are not in general equal.)
Although the value of the NMSE and its rate of decrease varied 
from image to image,
it generally either plateaued within a few iterations
or dropped to very low values (below $10^{-10}$).
In the latter case,
this suggests that the fixed point algorithm finds a $\dir$
such that $\dir+\model_\dir(\dir) = \tot$;
in the former case, we suspect there is no solution to the system
and we get $\dir+\model_\dir(\dir) \approx \tot$.
This behavior occurred for both local and global models.
We provide convergence plots in Figure~\ref{fig:convergence}.

\textbf{Noise.} In a separate, small experiment
(described in Section~\ref{sec:noise})
we explored how descattering accuracy was affected by noise.
Results showed that noise passed without amplification through the descattering process,
indicating that performing descattering does not necessitate additional noise removal beyond that performed in the reconstruction step (e.g., regularization).



\subsection{Oracle scatter fitting}
To explore the capacity of the models,
we evaluated them in an oracle fitting setting,
i.e.,
we evaluated their
ability to map from a single direct radiograph, $\dir$
to its corresponding scatter, $\sca$,
when both $\dir$ and $\sca$ are known
(provided by some oracle that one would not have in practice).
A model's performance at this task upper bounds
its descattering performance;
however,
it does not give a complete picture of a model's potential for descattering,
since models may overfit the training data and fail to generalize.

\renewcommand{\arraystretch}{.9}
\begin{table}[htbp]
    \centering
    \begin{tabular}{rrrrrrr}
    \toprule
     & \multicolumn{6}{c}{method} \\ \cmidrule{2-7}
    dataset   & w/out scatter & single field & parametric & multikernel & convolution & w/ scatter  \\ \midrule
    mono  & 0.017 & 0.017 & 0.168 & 0.061 & 0.192 & 1.268 \\
    poly & 0.268 & 0.268 & 0.358 & 0.371 & 0.364 & 1.281 \\
    \bottomrule
    \end{tabular}
    \caption{Maximum MADE (g/cm$^3$) after oracle scatter fitting.}
    \label{tab:oracle}
\end{table}


For each $(\dir_t, \sca_t)$ in the monoenergetic and polyenergetic testing sets,
we fit a \textbf{single field}, 
\textbf{parametric kernel}, \textbf{multi kernel}, and \textbf{convolutional} model.
(This is equivalent to performing local fitting with a single neighbor
if the training set includes the testing set.)
We used each fitted model to make a prediction, $\hat\sca_t$.
We formed an estimated direct transmission, $\hat{\dir}_t = \dir_t + \sca_t - \hat{\sca}_t$,
and reconstructed, yielding $\hat{\rho}_t$.
We compared these estimated densities
to the ground truth in terms of median absolute density error~\eqref{eq:MADE}.


The scatter fitting results (Table~\ref{tab:oracle})
show that all models have some potential to improve density reconstruction via descattering,
giving maximum density errors of under 0.2 g/cm$^3$
and 0.4 g/cm$^3$ in the monoenergetic and polyenergetic cases,
respectively,
as compared to errors of over 1.25 g/cm$^3$
when scatter is simply ignored.
The single field model performs nearly identically to
reconstruction without scatter;
this is expected, because each $\hat{\sca}_t$ and $\sca_t$
are identical except for the fact that,
as described in Section~\ref{sec:prelim},
we downsample by a factor of four during scatter fitting.
We can view the parametric model as a generalization of the convolutional model
with extra flexibility in the tuning of the parameters of the nonlinearity \eqref{eq:nonlinear}.
That the parametric model gives a smaller error than the convolutional model
here but not in the main experiment in Section~\ref{sec:exp:descatter} suggests that this extra flexibility
leads to some degree of overfitting, i.e., less generalization to training data.
The same observation holds for the multikernel model.

\subsection{Scatter estimation}
\label{sec:experiments:scatter}

We also assessed the methods' scatter estimation performance
without descattering involved, 
i.e., fitting on a training set
and predicting the corresponding $\sca$ for a given known $\dir$
from the testing set.
We assessed the results in terms of the median absolute density error \eqref{eq:MADE},
when reconstructing from $\tot - \model(\dir)$.

\begin{figure*}[!htb]
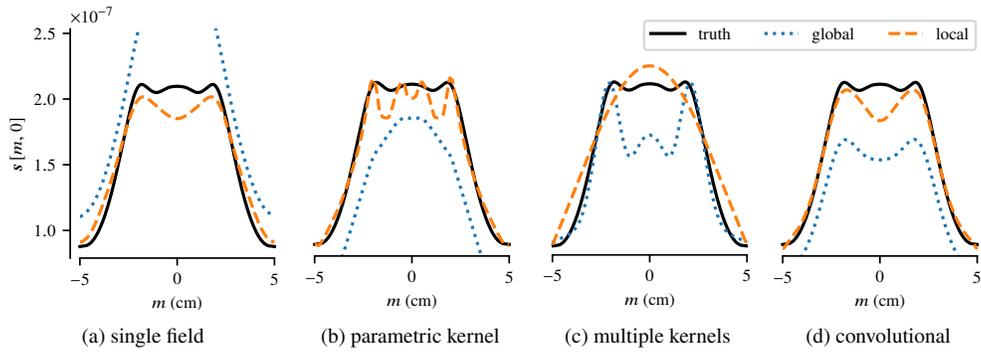

\captionsetup[subfigure]{justification=centering}
    \centering
    \begin{minipage}[b]{.29\linewidth}
    \begin{subfigure}[b]{\linewidth}
    \begin{adjustbox}{clip,trim=0cm .5cm .5cm .1cm, scale=.8}
    \input{figures/mono_scatter_estimation_from_d_single_field.pgf}%
    \end{adjustbox}
    \caption{single field}
    \end{subfigure}%
    \end{minipage}
    \begin{minipage}[b]{.23\linewidth}
    \begin{subfigure}[b]{\linewidth}
    \begin{adjustbox}{clip,trim=0.5cm .5cm 0cm 1cm, scale=.8}
    \input{figures/mono_scatter_estimation_from_d_parametric_kernel.pgf}%
    \end{adjustbox}
    \caption{parametric kernel}
    \end{subfigure}%
    \end{minipage}
    \begin{minipage}[b]{.46\linewidth}
    \begin{subfigure}{\linewidth}\hfill
    \begin{adjustbox}{clip,trim=.75cm .25cm .75cm .25cm, scale=.8}
\begingroup%
\makeatletter%
\begin{pgfpicture}%
\pgfpathrectangle{\pgfpointorigin}{\pgfqpoint{3.000000in}{0.500000in}}%
\pgfusepath{use as bounding box, clip}%
\begin{pgfscope}%
\pgfsetbuttcap%
\pgfsetmiterjoin%
\definecolor{currentfill}{rgb}{1.000000,1.000000,1.000000}%
\pgfsetfillcolor{currentfill}%
\pgfsetlinewidth{0.000000pt}%
\definecolor{currentstroke}{rgb}{1.000000,1.000000,1.000000}%
\pgfsetstrokecolor{currentstroke}%
\pgfsetdash{}{0pt}%
\pgfpathmoveto{\pgfqpoint{0.000000in}{0.000000in}}%
\pgfpathlineto{\pgfqpoint{3.000000in}{0.000000in}}%
\pgfpathlineto{\pgfqpoint{3.000000in}{0.500000in}}%
\pgfpathlineto{\pgfqpoint{0.000000in}{0.500000in}}%
\pgfpathclose%
\pgfusepath{fill}%
\end{pgfscope}%
\begin{pgfscope}%
\pgfsetbuttcap%
\pgfsetmiterjoin%
\definecolor{currentfill}{rgb}{1.000000,1.000000,1.000000}%
\pgfsetfillcolor{currentfill}%
\pgfsetfillopacity{0.800000}%
\pgfsetlinewidth{1.003750pt}%
\definecolor{currentstroke}{rgb}{0.800000,0.800000,0.800000}%
\pgfsetstrokecolor{currentstroke}%
\pgfsetstrokeopacity{0.800000}%
\pgfsetdash{}{0pt}%
\pgfpathmoveto{\pgfqpoint{0.437060in}{0.154639in}}%
\pgfpathlineto{\pgfqpoint{2.562940in}{0.154639in}}%
\pgfpathquadraticcurveto{\pgfqpoint{2.585162in}{0.154639in}}{\pgfqpoint{2.585162in}{0.176862in}}%
\pgfpathlineto{\pgfqpoint{2.585162in}{0.323138in}}%
\pgfpathquadraticcurveto{\pgfqpoint{2.585162in}{0.345361in}}{\pgfqpoint{2.562940in}{0.345361in}}%
\pgfpathlineto{\pgfqpoint{0.437060in}{0.345361in}}%
\pgfpathquadraticcurveto{\pgfqpoint{0.414838in}{0.345361in}}{\pgfqpoint{0.414838in}{0.323138in}}%
\pgfpathlineto{\pgfqpoint{0.414838in}{0.176862in}}%
\pgfpathquadraticcurveto{\pgfqpoint{0.414838in}{0.154639in}}{\pgfqpoint{0.437060in}{0.154639in}}%
\pgfpathclose%
\pgfusepath{stroke,fill}%
\end{pgfscope}%
\begin{pgfscope}%
\pgfsetrectcap%
\pgfsetroundjoin%
\pgfsetlinewidth{1.505625pt}%
\definecolor{currentstroke}{rgb}{0.000000,0.000000,0.000000}%
\pgfsetstrokecolor{currentstroke}%
\pgfsetdash{}{0pt}%
\pgfpathmoveto{\pgfqpoint{0.459282in}{0.262027in}}%
\pgfpathlineto{\pgfqpoint{0.681504in}{0.262027in}}%
\pgfusepath{stroke}%
\end{pgfscope}%
\begin{pgfscope}%
\definecolor{textcolor}{rgb}{0.000000,0.000000,0.000000}%
\pgfsetstrokecolor{textcolor}%
\pgfsetfillcolor{textcolor}%
\pgftext[x=0.770393in,y=0.223138in,left,base]{\color{textcolor}\fontsize{8.000000}{9.600000}\selectfont truth}%
\end{pgfscope}%
\begin{pgfscope}%
\pgfsetbuttcap%
\pgfsetroundjoin%
\pgfsetlinewidth{1.505625pt}%
\definecolor{currentstroke}{rgb}{0.121569,0.466667,0.705882}%
\pgfsetstrokecolor{currentstroke}%
\pgfsetdash{{1.500000pt}{2.475000pt}}{0.000000pt}%
\pgfpathmoveto{\pgfqpoint{1.202502in}{0.262027in}}%
\pgfpathlineto{\pgfqpoint{1.424724in}{0.262027in}}%
\pgfusepath{stroke}%
\end{pgfscope}%
\begin{pgfscope}%
\definecolor{textcolor}{rgb}{0.000000,0.000000,0.000000}%
\pgfsetstrokecolor{textcolor}%
\pgfsetfillcolor{textcolor}%
\pgftext[x=1.513613in,y=0.223138in,left,base]{\color{textcolor}\fontsize{8.000000}{9.600000}\selectfont global}%
\end{pgfscope}%
\begin{pgfscope}%
\pgfsetbuttcap%
\pgfsetroundjoin%
\pgfsetlinewidth{1.505625pt}%
\definecolor{currentstroke}{rgb}{1.000000,0.498039,0.054902}%
\pgfsetstrokecolor{currentstroke}%
\pgfsetdash{{5.550000pt}{2.400000pt}}{0.000000pt}%
\pgfpathmoveto{\pgfqpoint{2.013610in}{0.262027in}}%
\pgfpathlineto{\pgfqpoint{2.235832in}{0.262027in}}%
\pgfusepath{stroke}%
\end{pgfscope}%
\begin{pgfscope}%
\definecolor{textcolor}{rgb}{0.000000,0.000000,0.000000}%
\pgfsetstrokecolor{textcolor}%
\pgfsetfillcolor{textcolor}%
\pgftext[x=2.324721in,y=0.223138in,left,base]{\color{textcolor}\fontsize{8.000000}{9.600000}\selectfont local}%
\end{pgfscope}%
\end{pgfpicture}%
\makeatother%
\endgroup%
%
    \end{adjustbox}%
    \end{subfigure}\\
    \begin{subfigure}[b]{.5\linewidth}
    \begin{adjustbox}{clip,trim=.5cm .5cm 0cm 1.1cm, scale=.8}
    \input{figures/mono_scatter_estimation_from_d_three_kernels.pgf}%
    \end{adjustbox}%
    \caption{multiple kernels}
    \end{subfigure}%
    \begin{subfigure}[b]{.5\linewidth}
    \begin{adjustbox}{clip,trim=.5cm .5cm 0cm 1.1cm, scale=.8}
    \input{figures/mono_scatter_estimation_from_d_convolution.pgf}%
    \end{adjustbox}%
    \caption{convolutional}
    \end{subfigure}
    \end{minipage}
    \caption{Example scatter estimations from direct on the monoenergetic dataset,
    where fitting was performed on a separate training set.
        }
    \label{fig:scatter_estimation_profiles}
\end{figure*}

\begin{figure*}[!htb]
\captionsetup[subfigure]{justification=centering}
    \begin{subfigure}{.333\linewidth}
    \centering
        \includegraphics[width=\linewidth]{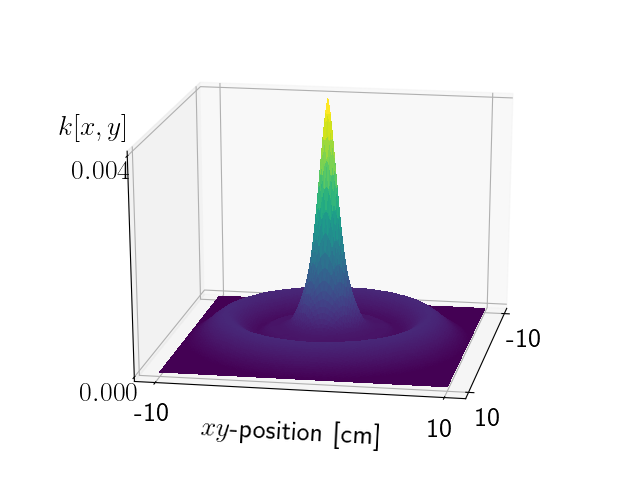}
        \caption{global kernel}
    \end{subfigure}%
    \begin{subfigure}{.333\linewidth}
    \centering
        \includegraphics[width=\linewidth]{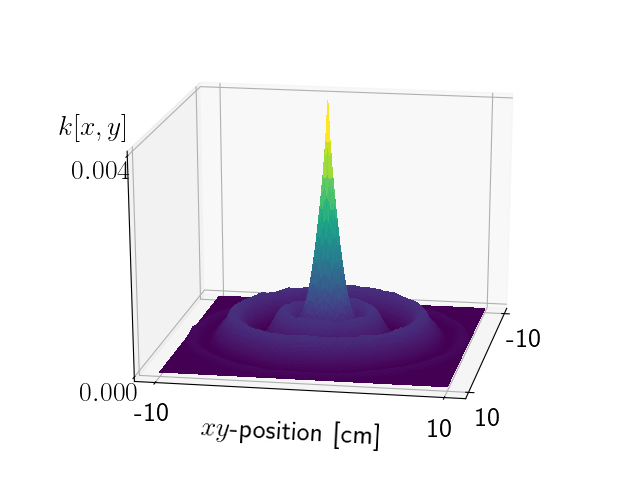}
        \caption{local kernel}
    \end{subfigure}%
    \begin{subfigure}{.333\linewidth}
    \centering
        \includegraphics[width=\linewidth]{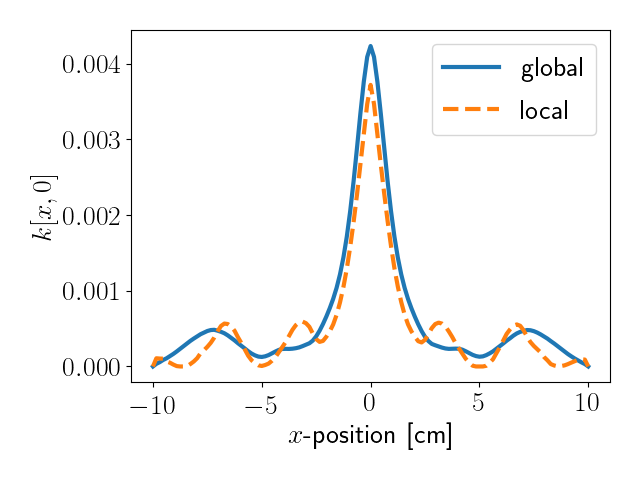}
    \caption{kernel profiles}
    \end{subfigure}%
    \caption{Qualitative comparison of the global kernel
    versus the local kernel learned for a single profile,
    plotted here as a function of position in pixels.
    The local kernels tend to be sharper.}
    \label{fig:kernels}
\end{figure*}

The results show that scatter estimation was not significantly easier 
than estimation and descattering (never more than 5\% lower error).
We interpret this to mean that the limiting factor for descattering performance is
accurately modeling scatter
as opposed to solving for $\dir$ from $\tot$.
We provide estimated scatter profiles in Figure~\ref{fig:scatter_estimation_profiles}
and example local and global kernels in Figure~\ref{fig:kernels}.

\subsection{Cases with increased scatter}
To investigate the performance of local descattering
as the amount of scatter increases,
we chose a single object from the polyenergetic dataset
and repeated the MCNP simulation with its mass scaled
by a factor of 1.5 and 2
(leaving the collimator unchanged).
These scaled objects have a much higher scatter-to-direct ratio
than the original set,
with a maximum ratio (over pixels) of 1.57 and 25.60, respectively,
compared to 0.34 in the original dataset.

\begin{figure}
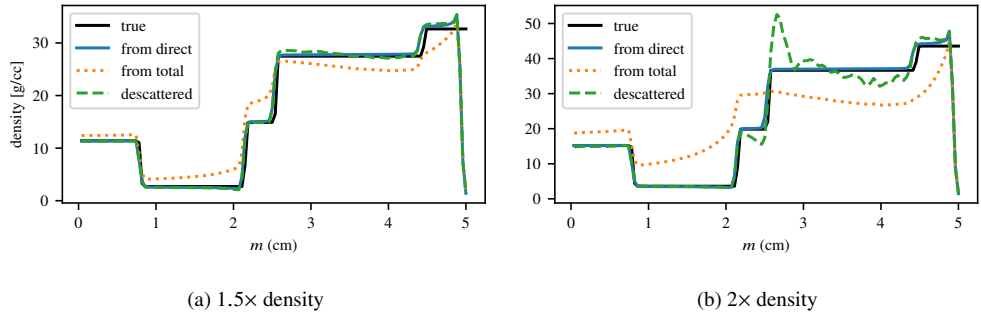

    \centering
    \begin{subfigure}{.5\linewidth}
    \centering
    \begin{adjustbox}{clip, trim=.3cm 0cm 0cm 0cm, scale=.75}
        \input{high_scatter_1pt5x.pgf}
        \end{adjustbox}
        \caption{$1.5\times$ density}
    \end{subfigure}%
    \begin{subfigure}{.5\linewidth}
    \centering
    \begin{adjustbox}{clip, trim=.62cm 0cm 0cm 0cm, scale=.75}
        \input{high_scatter_2x.pgf}
        \end{adjustbox}
        \caption{$2\times$ density}
    \end{subfigure}    
    \caption{Effect of increasing object density on reconstruction performance
    using a local convolutional model and fixed-point descattering (Algorithm~\ref{algo:fixed_point}).
    As the density scales up, the scatter-to-direct ratio increases,
    increasing the error.}
    \label{fig:high_scatter}
\end{figure}

For each of these scaled simulations,
we additionally perturbed the density by adding
or subtracting 1.0 or 2.0 g/cm$^3$ from the density of each shell,
creating four variations (-2.0, -1.0, +1.0, +2.0).
We fit a convolutional model to these four variations
and used it to descatter the scaled total transmission radiograph.
Thus, we can think of the four variations as forming a local training set.
We compare the original descattering result
to the two high-scatter cases in Figure~\ref{fig:high_scatter}.
The first trend we note is that,
as the density increases,
the reconstruction from the total radiograph (no descattering) deviates
more and more from the ground truth,
indicating that scatter has become a larger source of error.
While descattering is successful in the original simulation
and in the $1.5\times$ case,
it begins to struggle in the $2\times$ case,
reaching density errors of over 10 g/cm$^3$ in some regions.
However, 
the local approach is still better than ignoring scatter altogether.

\subsection{Effect of number of neighbors on reconstruction performance}
\label{sec:neighbors}
\begin{figure}
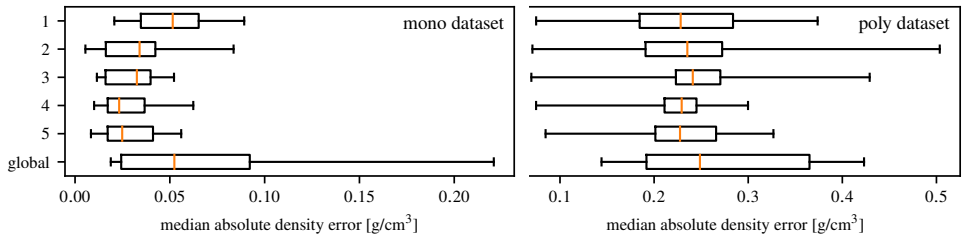

    \centering
    \begin{adjustbox}{clip, trim=.2cm 0cm .2cm 0cm, scale=0.8}
    \input{neighbors_mono.pgf}
    \end{adjustbox}
    \begin{adjustbox}{clip, trim=1.6cm 0cm .2cm 0cm, scale=0.8}
    \input{neighbors_Brem.pgf}
    \end{adjustbox}
    \caption{Effect of the neighborhood size on density reconstruction error
    using the local convolutional model.}
    \label{fig:neighbors}
\end{figure}

To evaluate the effect of the number of neighbors,
we repeated the descattering experiment with the local convolutional
model with the number of neighbors varying from one to five.
The results
(Figure~\ref{fig:neighbors})
show that the reconstruction error is
robust with respect to the number of neighbors,
with local fitting consistently outperforming
global no matter the number of neighbors.
Although we might expect that a small number of neighbors could lead to overfitting
and that increasing the number of neighbors may cause underfitting,
we do not see clear evidence of this here.
(Except that the global model, which is equivalent on our data to 
a local model with 90 neighbors,
performs worse than the local ones.)

\section{Discussion}
\label{sec:discussion}




\textbf{Local models vs CNNs.}
Our experiments showed that, using the proposed local fitting scheme,
the performance of simple models can rival that of a CNN (the DSE~\cite{maier_deep_2018});
we now compare and contrast these methods.
One advantage of the proposed method is that it does not require training a CNN,
which is slow and requires challenging hyperparameter tuning.
Another is its flexibility in terms of training:
in a setting where we know approximately what we want to image,
we could perform a few scatter simulations, fit a local convolutional model,
and then use it to descatter experimental data.
This aspect is especially appealing for applications wherein creating a training set with scatter of even tens of objects takes an unacceptable amount of time,
due to the complexity of the object or desired simulation accuracy.
Fitting a deep CNN on the same few scatter simulations would likely lead to overfitting.
It is also typically much easier to simulate the direct radiographs than to simulate scatter.
So in practice, we could simulate a dense set of direct radiographs for performing the local matching in our algorithm and simulate scatter only as necessary. The proposed local learning approach affords this flexibility.
The local fitting performed adaptively at reconstruction time also means that new examples can be added to the training dataset
and be used by the model without retraining.
Finally, the proposed model is simple to describe and easy to interpret,
in the sense that the learned kernels can be viewed as images
and intuitively encode the spatial component of scatter.
The main advantage of using a CNN over the proposed approach is that it is faster at testing time (milliseconds vs seconds to estimate a scatter radiograph).
This may be attractive in settings where data is collected quickly and scatter estimation needs to be immediate,
but may not matter in many practical settings.

Taking a broader perspective,
the local convolutional model is a piecewise linear
function from images to images, i.e., $\mathbb{R}^{N_1\times N_2} \to \mathbb{R}^{N_1 \times N_2}$,
with 
the pieces being defined by the neighborhoods
in the training data.
ReLU CNNs, including the DSE, are also piecewise linear functions~\cite{montufar_number_2014}.
The two models differ only in the parameterization
of the linear pieces and their boundaries.
These similarities may partly explain the similarity in performance between the two models.

\textbf{Limitations.}
Our simulations did not account for scatter within the detector,
which, depending on the hardware,
may contribute significantly to measured radiographs~\cite{bhatia_convolution_2017}.
Our local fitting method may be able to account for this type of scatter,
but it would need to be included in the training data.
Our testing datasets, although not small by the standards of other descattering work,
are probably not large enough to detect small differences between methods,
especially given the large amount of variation in the data.
Our experimental datasets consisted only of
spherically symmetric objects
and the performance of the scatter estimation may be different for different types of objects.
However, we note that spherical objects are themselves of practical importance in nuclear security applications
and, further,
there is nothing about the proposed scatter model that explicitly relies on this symmetry.
See Section~\ref{sec:ellipse}.
for a proof-of-concept experiment on an ellipsoidal object.
Finally,
in cases where the direct-to-scatter mapping is one-to-many,
no technique can deterministically estimate scatter from the direct radiograph perfectly,
including the proposed method.
However, the proposed technique may still be practically useful;
further, performance might be improved by finding neighbors based on total transmission, which could be unique even in a case where the direct signal is not.

\section{Conclusions and future work}
\label{sec:conclusion} 
Our experiments show that, when fit locally,
even simple models can describe scatter well,
thus avoiding the algorithmic challenges and computational burden of fitting and using a nonlinear or shift-variant model.
Of the four models we compared,
we recommend the local convolutional model,
given that it is easy to fit 
and provides excellent descattering performance.
The proposed approach offers performance on par with a recent deep learning-based
method,
with the advantage that it avoids an expensive network training step.

We see several possible extensions of this work.
In the context of X-ray CT,
our method can easily be used for multiview
(i.e., not spherically symmetric)
tomography problems, including medical imaging.
However, these settings may be challenging
if the variety in the objects to be imaged is large;
this would necessitate gathering a larger training set to ensure good neighbors can be found.
Validating the method in additional settings is an important next step.
We would also like to the study the effect of training set density 
on reconstruction performance;
ideally, we could find a rule of thumb for how many simulations
are needed to achieve a certain accuracy for a given application.
Another direction may be to study different distance metrics
for neighbor determination,
or to learn a good metric from the training data~\cite{kulis_metric_2013}.
Finally, we believe that the local modeling approach
may have applications outside of scatter estimation.
Wherever we have a training set of image pairs,
we can potentially fit simple, instance-adaptive local models
instead of complex, global models.

\paragraph{Acknowledgments.}
Erik Skau, Brendt Wohlberg, and Luke Hovey, LANL, provided useful discussions. 
Alexander Sietsema, Michigan State University, worked on related experiments and discussions.

\paragraph{Disclosures.}
The authors declare no conflicts of interest.

\paragraph{Data availability.}
Data underlying the results presented in this paper are not publicly available at this time but may be obtained from the authors upon reasonable request.

\paragraph{Supplemental document.}
See Supplement 1 for supporting content.

\bibliography{biblio_settings,references,refs_mike}



\pagebreak
\title{Local Models for Scatter Estimation and Descattering in Polyenergetic X-Ray Tomography: supplemental document}

\begin{abstract}
This document contains supplemental methods and figures for ``Local Models for Scatter Estimation and Descattering in Polyenergetic X-Ray Tomography.''
For the complete context,
refer to the main text,
Sections~\ref{sec:reconstruction} and \ref{sec:exp:descatter}.
\end{abstract}

\section{Reconstruction Algorithms}

\subsection{Monoenergetic Case} \label{sec:recon_mono}
In the monoenergetic, single-material case, 
the areal density from a given angle,
$\areal_v$,
can be easily recovered from the direct radiograph at that angle,
$\dir_v$.
Given a flat field image 
(an image with no object),
    $\flat[m,n] \approx C \photons_\text{in}(r_{m,n}),$
we have
$
    \frac{\dir_v[m,n]}{\flat[m,n]} = \exp(-\MAC \areal_v(r_{m,n})).
$
Taking the negative logarithm (pixelwise) and dividing by the mass attenuation coefficient gives $\areal_v(r)$.
Going from $\{\areal_v\}_{v=1}^{V}$ to $\dens$ is a standard tomography problem
that can be resolved by any of numerous direct~\cite{kak_principles_2001_nocity}, iterative~\cite{beister_iterative_2012}, or learning-based~\cite{mccann_biomedical_2019,ravishankar_image_2020} methods.

We also consider the monoenergetic, multimaterial case,
where there are nuisance objects:
elements of the scene that we are not interested in reconstructing.
If these can be imaged without the object of interest,
their effect can simply be included in the flat field radiograph
and divided out as described above.
If they cannot be imaged alone,
but we know the distribution of their mass,
we can compute their linear attenuation map 
(mass attenuation coefficient multiplied by areal density)
and subtract it after taking the $\log$.
Barring these types of prior information,
the reconstruction problem is much more challenging
and beyond the scope of this paper.

\subsection{Polyenergetic Case} \label{sec:recon_poly}
In the single-material, polyenergetic case,
we propose to follow the same two-step inversion as in the monoenergetic case;
however this is challenging 
because we cannot simply divide to remove the effect of the incident field $\photons_\text{in}$
(due to its dependence on energy)
and we cannot use a logarithm to invert the sum of exponentials that appears in \eqref{eq:direct-poly}.
To deal with $\photons_\text{in}$, 
we make the simplifying assumption that it is separable into 
an (unknown) space-dependant and a (known) energy-dependent term as 
 $\photons_\text{in}(r, E) = \photons_\text{in}(r) \photons_E(E)$
 with $\sum_e \photons_E(E_e) = 1$.
 Under this assumption, division by the flat field gives
 \begin{equation} \label{eq:poly-norm}
    \frac{\dir_v[m,n]}{\flat[m,n]} =
    \sum_e \photons_E(E_e)
    \exp \left(-\MAC(E_e) \areal_v(r_{m,n})\right).
 \end{equation}
Our solution relies on the fact that \eqref{eq:poly-norm}  is a monotone,
scalar function of $\areal_v$
and therefore invertible via a lookup table.
Our density reconstruction algorithm is as follows.
We consider \eqref{eq:poly-norm} as a scalar function of areal mass
\begin{equation} \label{eq:lookup-poly} 
    g(\areal) = \sum_e \photons_E(E_e)
     \exp \left(-\MAC(E_e) \areal \right).
\end{equation}
We use this expression to compute 
$g(\areal_0), g(\areal_1), \dots g(\areal_{L-1})$,
for a set of $L$ areal masses spaced evenly between 0 and some $\areal_{L-1} = \areal_\text{max}$,
the largest areal mass we expect to encounter
(if necessary, this can be determined iteratively).
For each pixel of the direct transmission,
we can then perform a lookup in $g$
to find the corresponding $\areal$
Once the $\{\areal_v\}$ are recovered,
reconstructing $\rho$ is a standard tomography problem,
exactly like in the monoenergetic case.

In the polyenergetic, multimaterial case,
we cannot treat the nuisance objects as part of the flat field and divide them out,
nor can we subtract off a known linear attenuation.
For a single nuisance object with known density field,
e.g., a collimator,
\begin{equation} \label{eq:lookup-poly-multi}
    g(\areal, m, n) = \\\sum_e \photons_E(E_e)
     \exp \left(- 
     \MAC(E_e) \areal - 
     \MAC_\eta(E_e) \areal_\eta(r_{m,n}) \right),
\end{equation}
where $\MAC_\eta(E_e)$ and $\areal_\eta$ denote
the known mass attenuation coefficient and areal mass of the nuisance object.
Inverting \eqref{eq:lookup-poly-multi} requires a three dimensional lookup table, 
which we can interpret as a version of the lookup table for \eqref{eq:lookup-poly}
for each location, $(m,n)$.
The same approach can be used for multiple nuisance objects
by adding more terms to the exponential in \eqref{eq:lookup-poly-multi}.

\section{Additional Results}
This section contains additional results from the experiments described in the main text,
Sections 4.3 and 4.5.
Please see these sections for further description.

\section{Descattering in the presence of noise} \label{sec:noise}
\begin{figure}
    \centering
    \input{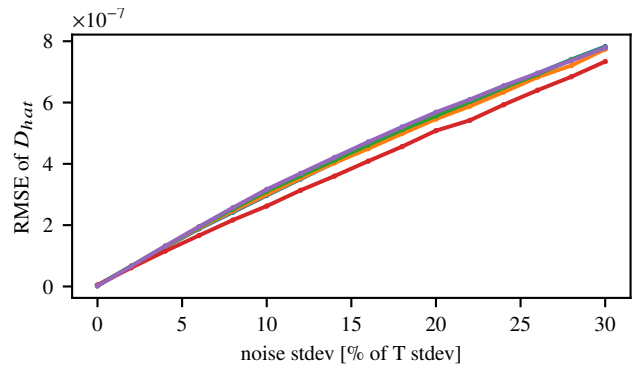}
    \caption{Noise after descattering as a function of noise in the total transmission.
    The relationship is linear, suggesting that noise passes without magnification through the descattering procedure.}
    \label{fig:noise}
\end{figure}

In this small experiment, we explored
whether the proposed descattering approach magnifies noise. 
 We added various levels of AWGN to the total transmission and performed descattering. 
 Our results (Figure~\ref{fig:noise}) show that noise passes though descattering without significant amplification, i.e., the RMSE of the descattered direct radiograph w.r.t the ground truth direct radiograph increases linearly with the standard deviation of the noise.
 
We also verified that, for the levels of noise in the above plot, the nearest neighbors of the total transmission radiographs are not changing. This explains the linear behavior with respect to the noise, because when the neighbors don’t change (with respect to same algorithm in noiseless case), the scatter model is a linear function. 

\section{Nonspherical objects} \label{sec:ellipse}
\begin{figure}
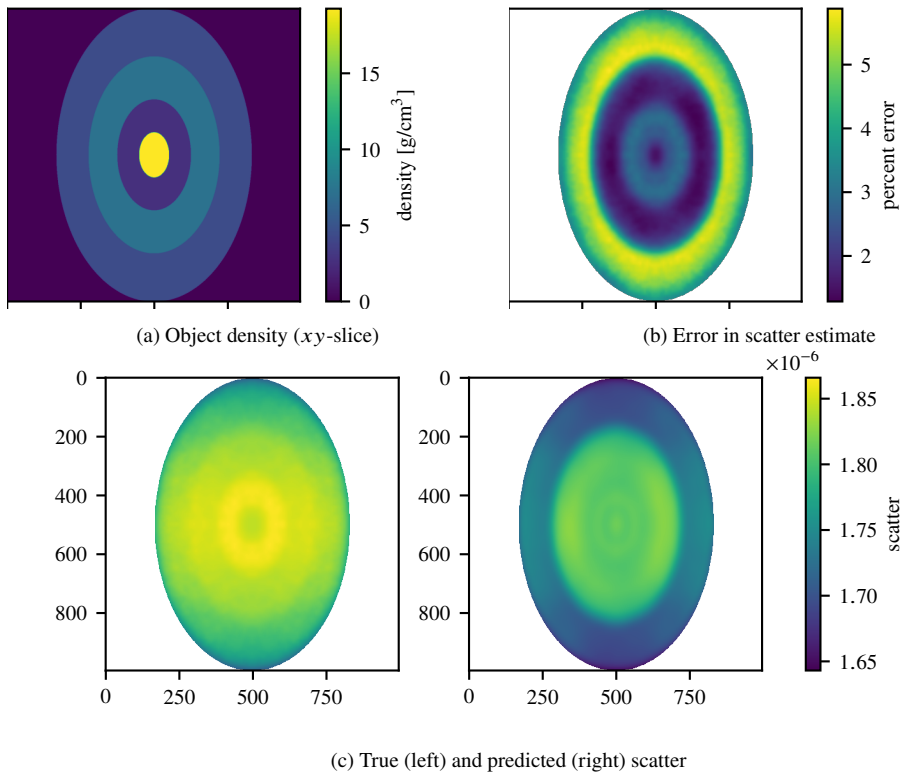

    \centering
    \begin{subfigure}{.5\linewidth}
    \begin{adjustbox}{clip,trim=2cm .75cm .1cm .25cm}
     \input{figures/rho_Elliptical_Elliptical.pgf}
    \end{adjustbox}
    \caption{Object density ($xy$-slice)}
    \end{subfigure}%
        \begin{subfigure}{.5\linewidth}
    \begin{adjustbox}{clip,trim=2cm .75cm .1cm .25cm}
    \input{figures/scatter_error_Elliptical_Elliptical.pgf}
    \end{adjustbox}
    \caption{Error in scatter estimate}
    \end{subfigure}\\
     \begin{subfigure}{\linewidth}
    \input{figures/scatter_Elliptical_Elliptical.pgf}
    \caption{True (left) and predicted (right) scatter}
    \end{subfigure}
    \caption{Proof-of-concept experiment on a nonspherical object.}
    \label{fig:ellipse}
\end{figure}

As a small proof-of-concept, we performed scatter estimation on a simulated ellipsoidal object using a set of four other ellipsoidal objects as training.
Results from the proposed local convolutional model are given in Figure~\ref{fig:ellipse}.
While preliminary, these results suggest that the proposed method can work on nonspherical objects.

\end{document}